\documentclass[]{imag-ms-template}

\title{Connectomics Informed by Large Language Models} 

\author{Elinor Thompson,$^{1,2\ast}$ Tiantian He,$^{1,3}$ Anna Schroder,$^{1,2}$ Ahmed Abdulaal,$^{1,3}$\\ Alec Sargood,$^{1,3}$ Sonja Soskic,$^{1,3}$ Henry F.J. Tregidgo,$^{1,2}$ Daniel C. Alexander$^{1,2\dagger}$\\\
{\small $^{1}$UCL Hawkes Institute, University College London, London, UK}\\
{\small $^{2}$Department of Computer Science, University College London, London, UK}\\
{\small $^{3}$Department of Medical Physics and Biomedical Engineering, University College London, London, UK}\\
{\small $^\ast$Correspondence:  elinor.thompson@ucl.ac.uk}\\
}

\begin{document} 

\maketitle 

\footnotetext[2]{%
Data used in preparation of this article were obtained from the Alzheimer's Disease Neuroimaging Initiative (ADNI) database (adni.loni.usc.edu). 
As such, the investigators within the ADNI contributed to the design and implementation of ADNI and/or provided data 
but did not participate in the analysis or writing of this report. 
A complete listing of ADNI investigators can be found at: 
\url{http://adni.loni.usc.edu/wp-content/uploads/how_to_apply/ADNI_Acknowledgement_List.pdf}.
}

\keywords{Tractography, Connectomics, Large Language Models}

\begin{abstract}
Tractography is a unique method for mapping white matter connections in the brain, but tractography algorithms suffer from an inherent trade-off between sensitivity and specificity that limits accuracy. Incorporating prior knowledge of white matter anatomy is an effective strategy for improving accuracy and has been successful for reducing false positives and false negatives in bundle-mapping protocols. However, it is challenging to scale this approach for connectomics due to the difficulty in synthesising information relating to many thousands of possible connections. In this work, we develop and evaluate a pipeline using large language models (LLMs) to generate quantitative priors for connectomics, based on their knowledge of neuroanatomy. We benchmark our approach against an evaluation set derived from a gold-standard tractography atlas, identifying prompting techniques to elicit accurate connectivity information from the LLMs. We further identify strategies for incorporating external knowledge sources into the pipeline, which can provide grounding for the LLM and improve accuracy. Finally, we demonstrate how the LLM-derived priors can augment existing tractography filtering approaches by identifying true-positive connections to retain during the filtering process. We show that these additional connections can improve the accuracy of a connectome-based model of pathology spread, which provides supporting evidence that the connections preserved by the LLM are valid.
\end{abstract}

\section{Introduction}

The human brain is a complex network, composed of functional units of grey matter connected by white matter fibres. A comprehensive map of these elements and their interconnections is termed the ``connectome''~\citep{sporns_human_2005}. Understanding this wiring diagram is critical to improve our understanding of brain function in health and disease~\citep{lappalainen_connectome-constrained_2024,saygin_anatomical_2012,passingham_anatomical_2002}. Furthermore, accurate characterisation of the connectome is important for applications such as: modelling the spread of neurodegenerative pathology along white matter connections~\citep{vogel_connectome-based_2023}; the identification of targets for neuromodulation~\citep{horn_connectivity_2017,sadeghzadeh_emerging_2024}; and neurosurgical planning~\citep{kamagata_advancements_2024}. 

Tractography is a technique for mapping the connectome that uses local fibre orientation estimates from diffusion MRI (dMRI) to trace the pathways of white matter bundles~\citep{jbabdi_tractography_2011,jeurissen_diffusion_2019}. It is the only technique that allows us to probe the structural connectivity of the brain non-invasively. This provides insight into inter-individual variability that is not possible with invasive methods. However, dMRI is a noisy and indirect measure of the underlying fibre architecture, leading to the reconstruction of spurious pathways (false positives), as well as missing connections (false negatives)~\citep{maier-hein_challenge_2017}. Tractography suffers from an inherent trade-off where increased specificity comes at the cost of reduced sensitivity, and vice-versa~\citep{schilling_brain_2020,zalesky_connectome_2016}.

Incorporating different sources of information into the tractography process can help to overcome this trade-off and trace valid connections without sacrificing accuracy~\citep{schilling_brain_2020}. Many tractogram filtering techniques use microstructural modelling of the dMRI signal to inform removal of false positives~\citep{smith_sift2_2015,daducci_commit_2015,schiavi_new_2020}. However, true positive connections can also be removed during filtering~\citep{sarwar_evaluation_2023}. A strategy to improve both sensitivity and specificity is to incorporate anatomical priors into the tracking process. This is the basis of most methods for ``virtual dissection'' of fibre bundles~\citep{rheault_bundle-specific_2019,warrington_xtract_2020,yendiki_automated_2011}, in which the priors are typically regions-of-interest (ROIs) that constrain tractography based on information about the bundle trajectory. Ground-truth knowledge for these anatomical priors comes from invasive methods, such as post-mortem dissection~\citep{catani_connectomic_2013} and tracer studies~\citep{morecraft_classic_2013}. These are time consuming and costly, which means that our neuroanatomical knowledge is still patchy and incomplete. Furthermore, different brain regions and white matter bundles are inconsistently described among texts, which makes it challenging to leverage this information for tractography in a consistent and reliable way~\citep{dulyan_navigating_2025,mandonnet_nomenclature_2018}. Although applying anatomical priors is effective when mapping individual bundles, it is not a feasible to scale this approach to a connectome comprising many thousands of potential edges, and encompassing both the large, long-range fibre bundles as well as the short u-fibres that connect neighbouring gyri. 

Here, we propose the use of Large Language Models (LLMs) to generate anatomical priors for tractography. LLMs have recently emerged as powerful tools for synthesising large amounts of textual data, and have gained popularity for their abilities in natural language processing tasks~\citep{brown_language_2020}. LLMs are typically based on a transformer architecture, and are trained on large amounts of text, including scientific papers. They should therefore contain substantial information about the connectivity architecture of the brain that should be useful to inform connectome construction with tractography. In this work, we build an automated pipeline using LLMs that can generate priors for each potential edge in the connectome, and is flexible to different parcellation schemes. Although the knowledge learned by the LLMs will be incomplete and imperfect, we test the potential of this idea to reduce false positives and false negatives in the standard connectome-construction pipeline. We propose and compare various prompting strategies and benchmark the performance of different models against an evaluation set, which is derived from a gold-standard tractography atlas. We also show how the LLM-informed priors can be augmented with specialist knowledge, for example from scientific papers, through information retrieval techniques. We show that this is particularly beneficial in the case where the parcellation contains regions that are not well known by the LLM. Finally, we incorporate LLM-derived priors into the tractography filtering process to help to retain valid connections that would otherwise be removed, and demonstrate the value of this approach in a model of pathology spread between connected brain regions.

\subsection{Related work: Using neuroanatomical knowledge to inform tractography}

Broadly speaking, tractography applications fall into two categories: bundle dissection and connectomics. Bundle dissection methods use anatomical priors to constrain tractography and retain streamlines belonging to a particular bundle. These constraints are typically applied in the form of ROIs that indicate the start and end points of the bundle, as well as regions that should be included or excluded from its trajectory. These can be drawn manually or derived from an atlas. Constraints can be used post-hoc to select the appropriate streamlines from a tractogram~\citep{wassermann_white_2016}, or used during the bundle generation process~\citep{warrington_xtract_2020,catani_virtual_2002,yendiki_automated_2011,chamberland_active_2017}. \citet{schilling_brain_2020} showed that using prior information in this way can significantly improve the accuracy of estimated brain connections, compared to unconstrained tractography. However, there is significant variability between different methods, partly due to inconsistency in the nomenclature and definitions of white matter bundles~\citep{schilling_tractography_2021,rheault_tractostorm_2020}.

In contrast, connectome generation typically involves generating streamlines across the white matter without explicit anatomical constraints. Connection strengths between region pairs are then quantified according to e.g. the number of streamlines connecting them, or microstructural measures along the pathway. Post-hoc filtering can remove false-positive streamlines, using microstructural information from the dMRI signal~\citep{smith_sift2_2015,daducci_commit_2015}, or heuristics based on streamline geometry~\citep{petit_structural_2023,astolfi_supervised_2023}, but these methods do not leverage neuroanatomical knowledge on the likelihood of a pathway existing between specific brain region pairs. In this work we use LLMs to provide a novel route for injecting this information into the filtering process, which is the first demonstration of neuroanatomical priors at the region-to-region connectome level rather than for individual bundles.

\section{Methods}
\label{sec:methods}
\begin{figure}[htbp]\begin{center}
  \includegraphics[width=\textwidth]{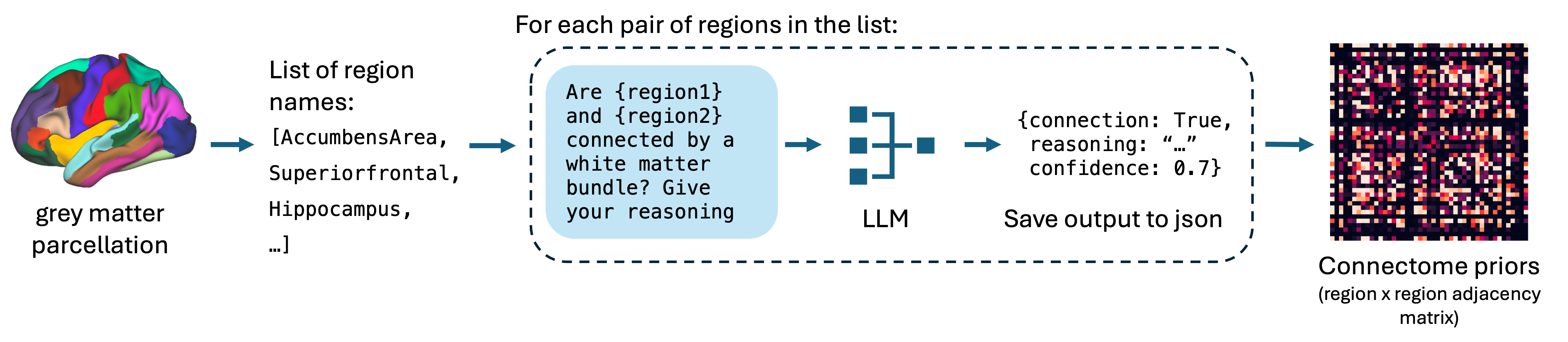}
  \caption{A schematic overview of the pipeline. The inputs are region names from a grey matter parcellation, and for each pair of regions the LLM is queried on the likelihood of a white matter connection between them. The outputs are saved in a machine-readable format that can be used to generate quantitative priors.}
  \label{fig:methods_overview}
  \end{center}
\end{figure}

We use LLMs to generate connectome priors by iteratively querying the existence of a white matter connection between pairs of brain regions, as shown in Figure \ref{fig:methods_overview}. The input into the pipeline is a list of the regions from a parcellation, which must have anatomically recognisable names (see section \ref{sec:parcellations} for details of the parcellations used in this study). For each unique pair of regions in this list, we query the existence of a white matter bundle connecting them. We explore different prompt engineering approaches for this query in section \ref{sec:prompt_engineering}. The outputs for each region pair are saved in json files, containing any reasoning output from the model and a final True or False output indicating whether the model classifies the regions as connected. The log probability assigned to this True or False token is used to give a model confidence score, where available. Log probabilities of tokens indicate the likelihood of each token occurring given the context. The jsons can then be read using a Python script to assign a prior for each connectome edge, which can be binary or weighted by the model's confidence.

\subsection{Evaluating LLM outputs}\label{sec:eval}
We benchmarked the LLMs against a well-known tractography atlas, the O'Donnell Research Group (ORG) white matter atlas \citep{zhang_anatomically_2018}. This atlas is based on tractography data from 100 individuals from the Human Connectome Project, which were combined into a single tractogram and clustered to form 800 bundles. The bundles were annotated using the White Matter Query Language~\citep{wassermann_white_2016} and manually inspected to remove false positives. We used this atlas for our evaluation set because it has undergone assessment by neuroanatomists, whilst encompassing both the long-range bundles and short u-fibres, which are missing from many other atlases. We combined the atlas clusters into a single reference tractogram, which was then used to generate reference connectomes for each parcellation. Streamline counts between regions were used as the connectivity metric. To make the evaluation set, we used the 50 most strongly connected within-hemisphere region pairs, and randomly selected 50 of the region pairs with no streamlines connecting them. This provided a balanced evaluation set of connected and unconnected pairs to test sensitivity and specificity. We decided on an evaluation set of 100 pairs as this tests a wide range of anatomical knowledge for the LLMs, while keeping costs low enough for multiple evaluations. For evaluation, we focus on within-hemisphere connections, assuming that these are broadly equivalent across right and left hemispheres. 

\subsection{Prompt engineering}\label{sec:prompt_engineering}
We compared three different prompting approaches, with increasing levels of reasoning required from the LLM:

\begin{itemize}
    \item \textit{minimal prompt}: the LLM is queried on the presence of a white matter connection existing between the two regions, and returns its answer as a Boolean.
    \item \textit{reasoning prompt}: the LLM is asked to provide a long-form text answer to the query from the minimal prompt, before summarising as True or False.
    \item \textit{chain-of-thought (COT) prompt}: the LLM is first asked to consider each of the relevant brain regions and their white matter connectivity patterns, before continuing to the reasoning prompt.
\end{itemize} 

The full text of these prompts is given in Figure \ref{fig:prompts}. In each case, a confidence score is derived from the final True or False token output by the LLM. The prompts for each region pair were processed in isolation, ensuring that prior interactions did not influence the results. However, for the reasoning and chain-of-thought prompts, the conversation history was preserved within a single session so that the LLM could access the previous reasoning steps in the conversation. LLM outputs are stochastic, so we performed repeats to evaluate reliability. Furthermore, the order that information is presented in the prompt is known to affect the accuracy of the LLM's response, based on common patterns in the training data (i.e. an LLM trained on ``A is B'' will not necessarily generalise to ``B is A''~\citep{berglund_reversal_2024}), so we repeated each experiment with the region pairs presented in both possible configurations (i.e. ``Is region A connected to region B'' and ``Is region B connected to region A'').

We also tested an approach where the LLM is explicitly encouraged to admit uncertainty (i.e. ``if you don't know, write `don't know'\,''), as this is thought to reduce hallucinations~\citep{zhou_context-faithful_2023}. We refer to this as the ``uncertainty prompt variant'' (UPV). When the LLM expresses uncertainty about the existence of a connection, we record the output as ``False'', treating the region pair as not connected. This is based on the assumption that the LLM is likely to be uncertain about connections that are not documented in the literature. Since the training data for LLMs typically lacks negative examples, such as explicit statements confirming the absence of a connection, uncertainty in the model's output may reflect this lack of evidence. We compared the performance of the three original prompts with and without the uncertainty variant, leading to six prompting strategies in total.

\subsection{Retrieval Augmented Generation}\label{sec:RAG_methods}

Retrieval augmented generation (RAG) describes the process of supplementing the LLM's knowledge with external documents~\citep{lewis_retrieval-augmented_2021}. Given a query, RAG aims to the retrieve the most relevant information to provide additional context, which is then shown to the LLM as part of the prompt. We demonstrate the use of RAG in two contexts: providing information about a parcellation where the regions are not present in the LLM's training corpus; and providing citations for the brain connections identified by the LLM.

\subsubsection{Using RAG to provide additional context where the LLM lacks knowledge about the parcellation}
\begin{figure}[htbp]
  \begin{center}

  \includegraphics[width=0.8\textwidth]{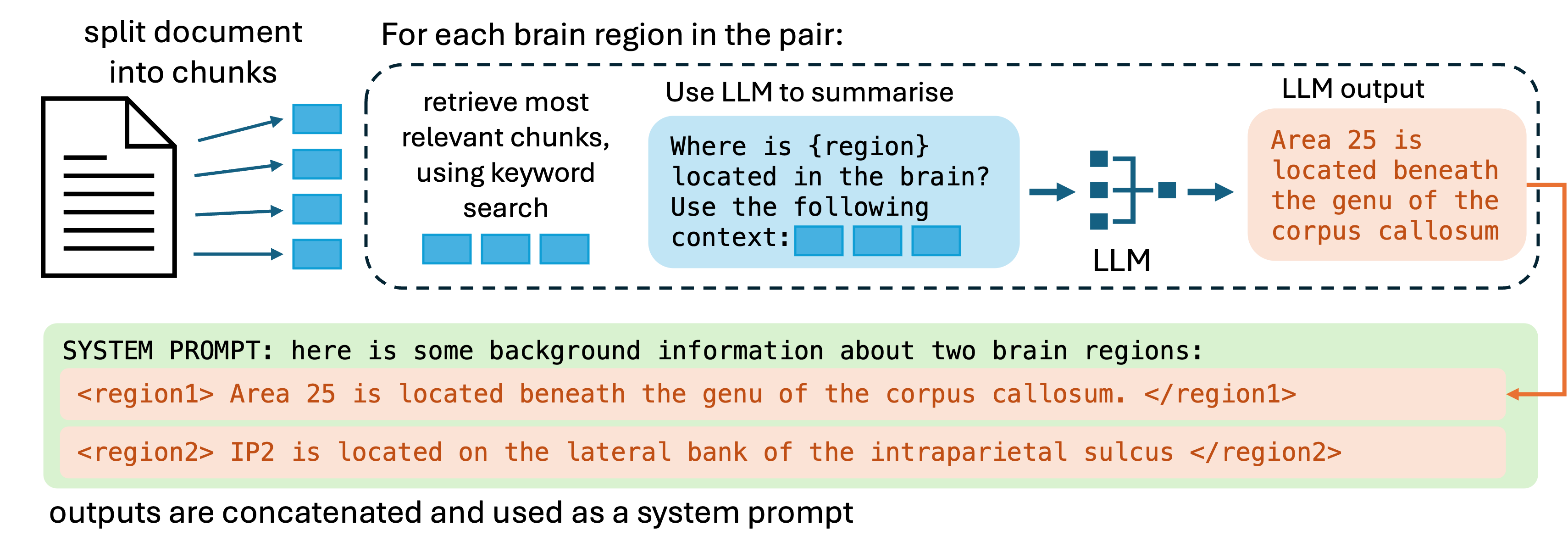}
  \caption{Schematic diagram showing the RAG pipeline for providing the LLM with contextual information about the parcellation. For each brain region, we retrieve relevant chunks from the document using keyword search, and pass these to the LLM to generate a short summary describing the location of the region. These summaries are then included in a system prompt.}
  \label{fig:RAG_parcellation}
\end{center}
\end{figure}

The RAG pipeline is shown in Figure \ref{fig:RAG_parcellation}. The aim is to provide the LLM with additional information about the regions in the parcellation. We split the document using Langchain's semantic chunking tool, which groups sentences with similar embeddings, and used BM25~\citep{robertson_probabilistic_2009} keyword search to find the three most relevant text chunks for each region. The LLM was then tasked with using this information to summarise the location of the brain region in question. The LLM's summaries for each region in a pair were concatenated and used in a system prompt, which is prepended to the prompting approaches described in section \ref{sec:prompt_engineering}. System prompts are prompts that start a conversation, which provide overarching instructions and context to the LLM. 

\subsubsection{Using RAG to ground responses in verifiable sources}\label{sec:RAG_connect_methods}
\begin{figure}[htbp]
  \begin{center}
  \includegraphics[width=1\textwidth]{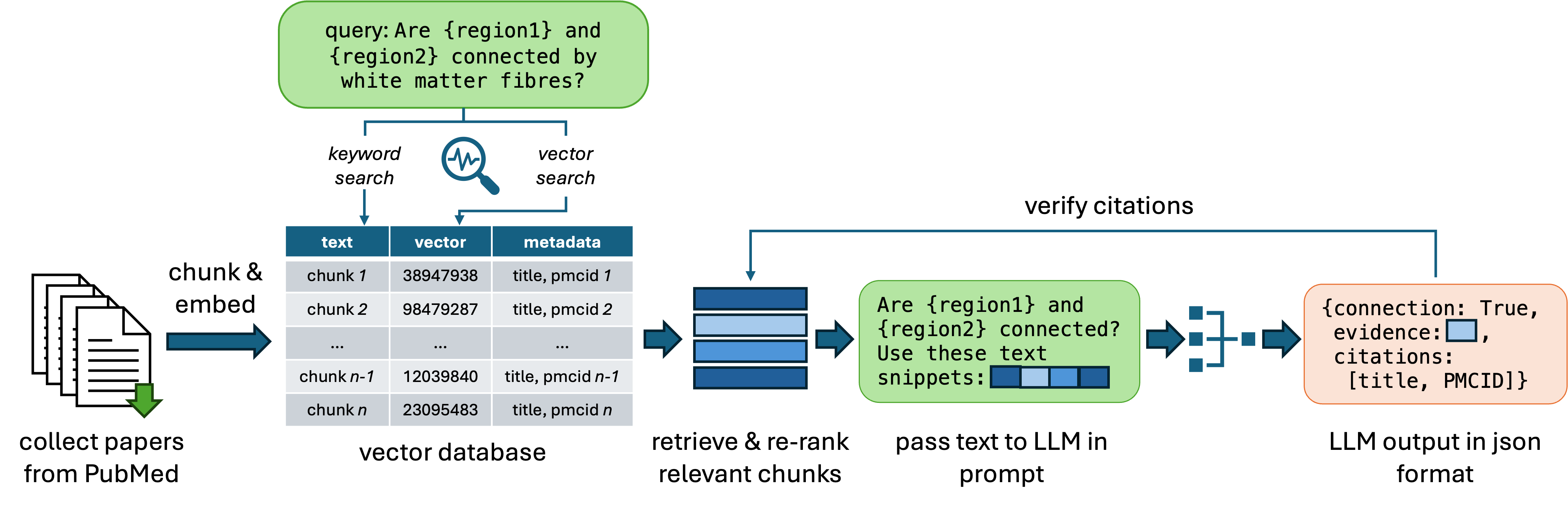}
  \caption{Schematic diagram showing the RAG pipeline for grounding the LLM's responses in texts from a database of neuroscience papers. Relevant papers were downloaded from PubMed and split into chunks. The chunks were stored in a database alongside their vector embeddings and article metadata. Hybrid search was used to retrieve relevant chunks, which were shown to the LLM in the prompt to use in its classification of the region pair.}
  \label{fig:RAG_connect}
  \end{center}
\end{figure}

We further used RAG as a way of grounding the LLM's responses in verifiable citations from a database of published neuroscience literature, as shown in Figure \ref{fig:RAG_connect}. To generate the database, we downloaded open-access papers in bulk from PubMed using the pubget Python package (\url{https://github.com/neuroquery/pubget}). We used a search query to confine our search to relevant papers containing any of the following terms in their abstract: ``white matter'', ``tractography'', or ``structural connectivity'', yielding 20,160 publications. The full text of the papers were downloaded and concatenated with their abstract texts, then split into chunks using Langchain's recursive text splitter. We used chunks of 2500 characters each, with 200 character chunk overlap, as this has been shown to be an optimal chunk size for RAG~\citep{wang_searching_2024}. The resultant 360,549 text chunks were then converted into numerical vectors encoding their semantic meaning, using a pre-trained embedding model, Cohere's `embed-english-v3.0', which has been designed especially for RAG applications. The chunks and their corresponding vectors were stored in a vector database alongside the corresponding article title, PubMed keywords, and PubMed Central Unique ID (PMCID), using LanceDB (\url{https://lancedb.com}).

The next step is to retrieve relevant text chunks that can help the LLM to ascertain whether a connection exists between the relevant brain regions. We used LanceDB's hybrid search, which combines vector similarity with keyword search to improve fidelity of the retrieval~\citep{sawarkar_blended_2024}, since the brain regions might be described differently in different texts. Cohere's reranking model (`rerank-english-v3.0') was then used to improve the search results by reordering the 20 highest-scoring chunks according to semantic relevance to the query. The highest scoring five chunks after re-ranking were shown to the LLM in the prompt, to inform its response. We then prompt the LLM to use the texts to determine whether the brain regions under consideration are connected by white matter pathways, and output the text snippets that support its response alongside the corresponding article title and PMCID. The full text of the prompt is given in the Appendix. There is a final verification step to check that the titles and the PMCIDs output by the LLM correspond to valid entries from the database.

\subsection{LLM-augmented filtering}\label{sec:filtering_methods}
We augmented a standard tractography filtering method with LLM-derived priors. This approach was designed to complement the imaging-based information used by the filtering algorithm with the LLM's knowledge of neuroanatomy, to improve the sensitivity of the technique. Filtering was performed on tractograms generated using probabilistic tractography, with the additional step that connections that were classified as True by the LLM were preserved even if they would otherwise be removed during filtering. Conversely, any connection surviving the filtering step was retained irrespective of the LLM classification. In this way, the tractogram remains the primary source of evidence, with the LLM serving to safeguard against the loss of true positive connections. The LLM does not remove existing connections or introduce new ones that do not already exist in the tractogram.

\section{Experiments}\label{experiments}

\subsection{Parcellations}\label{sec:parcellations}
Firstly, we used the Desikan-Killiany-Tourville (DKT) parcellation~\citep{klein_101_2012,manera_cerebra_2020}, which is commonly used in neuroimaging studies. The parcellation contains 45 grey matter parcels in each of the left and right hemispheres, delineated according to landmarks in the sulcal fundi and the subcortical grey matter.

We also used the Human Connectome Project's Multi-Modal Parcellation (HCP-MMP)~\citep{glasser_multi-modal_2016}. This is a more fine-grained parcellation, comprising 180 regions in each hemisphere of the cortical grey matter. The regions were delineated using a semi-automated approach, with their boundaries following sharp changes in multi-modal cortical properties. Over half of these regions were newly described in the HCP-MMP, which provides a challenging scenario for the LLM since the region names will not feature heavily in its training corpus.

\subsection{Comparing Prompt Engineering Approaches}
\begin{figure}[htbp]
  \begin{center}
  \includegraphics[width=\textwidth]{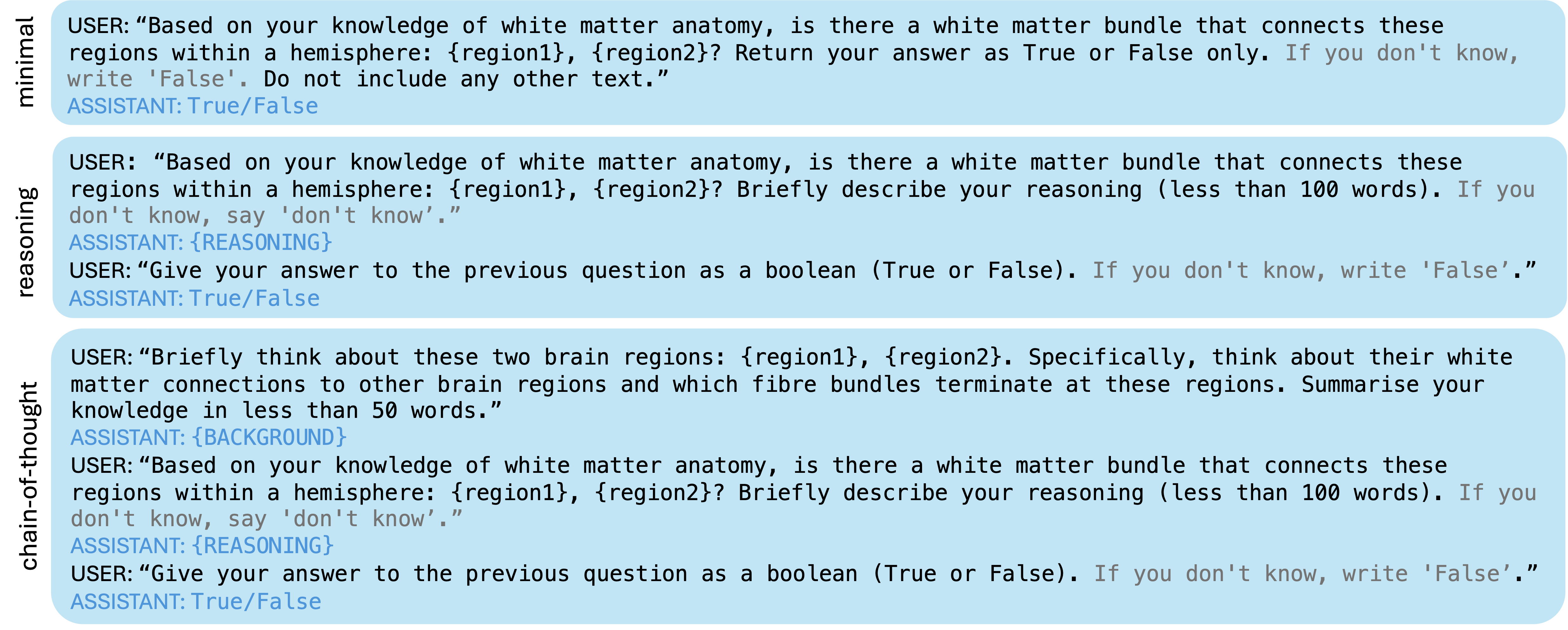}
  \caption{The text of the prompts used. The optional uncertainty prompt variant (UPV) is shown in grey font.}
  \label{fig:prompts}
  \end{center}
\end{figure}

We evaluated the six different prompting strategies described in section \ref{sec:prompt_engineering} (Figure \ref{fig:prompts}), testing each across four widely used language models: Claude 3.5 Sonnet (Anthropic), Llama 3 (8B, Meta), GPT-4 Turbo (OpenAI), and GPT-4o (OpenAI). Llama 3 was accessed via the Groq API, while the remaining models were accessed through their respective developers’ APIs. To improve efficiency, we employed asynchronous API calls, enabling parallel processing across region pairs.

We benchmarked the results against the evaluation set described in section \ref{sec:eval}, with the DKT parcellation. The experiment was repeated four times with each prompt, with the ordering of the regions in the prompt switched after two repeats. Log-probabilities assigned to the final classification token were used to calculate model confidence scores.

\subsection{Evaluating the RAG pipeline}
\subsubsection{Providing context about parcellation regions.}
Preliminary experiments showed that the LLM was lacking knowledge about many regions in the HCP-MMP, so we tested whether supplying additional information to the model about the parcellation could improve its ability to identify connections. We used the 80-page supplementary information from the HCP-MMP publication as the text corpus~\citep{glasser_multi-modal_2016}, split into 332 semantically similar chunks. We evaluated the performance of GPT-4 Turbo on the HCP-MMP evaluation set, with and without the RAG pipeline described in section \ref{sec:RAG_methods}. We used the minimal and COT prompts, since they had performed best for the previous experiment. We repeated the experiment twice with each ordering of the input region pairs and each prompt.

\subsubsection{Retrieving evidence for connections from neuroscience literature.} We developed a RAG pipeline to retrieve relevant text snippets from a database of neuroscience papers, to help the LLM classify region pairs as connected or unconnected, as detailed in section \ref{sec:RAG_connect_methods}. Using GPT-4 Turbo, the pipeline was tested on the evaluation set from the DKT parcellation. We repeated the experiment twice with each ordering of the input region pairs.

\subsection{Evaluating LLM-augmented filtering with a network spreading model of pathology propagation}
We evaluated the accuracy of LLM-augmented filtering by using the filtered connectomes as substrates in a network-based diffusion model of pathology spread~\citep{raj_network_2012}. These models simulate the spread of pathology through the brain network, reflecting the transneuronal spread of pathological proteins in neurodegenerative disease~\citep{vogel_connectome-based_2023}. One such protein is tau, which becomes misfolded in Alzheimer's disease to form neurofibrillary tangles, and can be measured in-vivo using tau-PET~\citep{groot_tau_2022}. Previous work has shown that network spreading models provide a good fit to patterns of tau deposition in Alzheimer's~\citep{schafer_network_2020,raj_combined_2021,vogel_spread_2020}, supporting the hypothesis that misfolded tau spreads along white matter fibres.

\subsubsection{Tractography and connectome filtering}
First, we performed whole-brain tractography on 50 unrelated subjects from the Human Connectome Project (HCP)~\citep{van_essen_wu-minn_2013}, using MRtrix3 tools~\citep{tournier_mrtrix3_2019}. Fibre orientations were obtained using multi-shell multi-tissue constrained spherical deconvolution~\citep{jeurissen_multi-tissue_2014}. We then performed probabilistic tracking using the iFOD2 algorithm to obtain five million streamlines for each subject. Anatomically constrained tractography (ACT) was employed to improve biological plausibility~\citep{smith_anatomically-constrained_2012}, seeding from a white matter mask. 

COMMIT2, a commonly used microstructure-informed tractography filtering algorithm, was used as our baseline method~\citep{daducci_commit_2015,schiavi_new_2020}. COMMIT2 outputs a weight for each streamline that represents its contribution to the measured diffusion signal. Typically, connections are removed if the COMMIT2 weights between two ROIs sum to zero. We used a regularisation parameter of $\lambda = 5 \times 10^{-4}$, as this performed best in preliminary testing across a wider range of values ($5 \times 10^{-5}$ to $5 \times 10^{-2})$.

Anatomical priors were generated using the minimal prompt with GPT-4 Turbo, as this combination had shown good balance between cost and accuracy in previous experiments (section \ref{sec:prompt_engineering_results}). The prompt from figure \ref{fig:prompts} was adjusted to account for both inter- and intra-hemispheric connections. The priors were calculated as the average model confidence scores across two repeats, once with each ordering of regions in the prompt. We used a confidence cut-off of 0.5, so that all connections classified as ``connected'' by the LLM were retained. Thus, connections were retained if either the COMMIT2 weights were greater than zero \textit{or} the LLM classified them as connected, and connections with both zero COMMIT2 weighting \textit{and} low LLM confidence were removed. 

\subsubsection{Fitting the network diffusion model}
The network diffusion model (NDM)~\citep{raj_network_2012} was used to demonstrate the benefit of the LLM-augmented filtering approach. The model simulates the spread of pathology through the brain network. Pathology is initiated at a seed region and propagated from this starting point along brain connections. It is typical to use a template healthy connectome as a substrate for the model, due to lack of good quality connectivity data in Alzheimer's patients and the long time-frames over which tau spread occurs. We fit the NDM to a group-average tau-PET pattern from individuals on the Alzheimer's disease continuum and compare the model fit across three connectomes for each HCP individual: unfiltered, COMMIT2 filtered, LLM-augmented-COMMIT2 filtered. An improvement in model fit suggests the LLM-derived priors improve overall accuracy of the connectome. As a null comparison, we performed a permutation test whereby for each connectome we measured the number of additional connections, $N$, that were retained when using the LLM-derived priors, compared to COMMIT2 filtering alone. We then added $N$ randomly-chosen connections to the COMMIT2-filtered connectome, and repeated the evaluation with the network diffusion model. This process was repeated 1000 times for each connectome, to allow us to test whether the connections identified by the LLM provided a significant improvement to model fit (and therefore connectome accuracy), compared to randomly chosen connections.

We used tau-PET data from the Alzheimer’s Disease Neuroimaging Initiative (ADNI) database \\(\href{https://adni.loni.usc.edu}{adni.loni.usc.edu}). The ADNI was launched in 2003 as a public-private partnership, led by Principal Investigator Michael W. Weiner, MD. The primary goal of ADNI has been to test whether serial magnetic resonance imaging (MRI), positron emission tomography (PET), other biological markers, and clinical and neuropsychological assessment can be combined to measure the progression of mild cognitive impairment (MCI) and early Alzheimer’s disease (AD). For up-to-date information on the ADNI, see \href{www.adni-info.org}{www.adni-info.org}. We evaluated the model fit to averaged tau-PET data from 242 amyloid-positive, tau-positive individuals, with age at scan 77.2 $\pm$ 8.4 years. Tau positivity was defined using Gaussian mixture modelling. We used regional flortaucipir standardised uptake value ratios (SUVRs) that had been pre-processed according to the \href{https://adni.bitbucket.io/reference/docs/UCBERKELEYAV1451/UCBERKELEY_AV1451_Methods_Aug2018.pdf}{ADNI Flortaucipir processing pipeline}, for each region in the Desikan-Killiany atlas~\citep{desikan_automated_2006}. Subcortical regions were not used to fit the model due to known issues with off-target binding of flortaucipir in the choroid plexus that can affect these areas~\citep{groot_tau_2022}.

We compared the model fit across each of the binarised networks, using our group's software toolbox~\citep{thompson_demonstration_2024}. Model fit was evaluated by the Pearson's correlation (r), and the sum-of-squared errors (SSE) between the model prediction and the measured tau-PET data. 

\section{Results}
\subsection{Prompt Engineering}\label{sec:prompt_engineering_results}
Table \ref{table:prompt_accuracy} shows the accuracy of the different prompting strategies and models in the evaluation set of 50 connected region pairs and 50 unconnected pairs from the DKT parcellation. Accuracy was measured as the number of region pairs correctly classified by the LLM as connected or not connected, as a percentage of the total evaluation set of 100 pairs. The highest accuracy, 91\%, was achieved by GPT-4 Turbo with the uncertainty variant of the COT prompt. All models achieved at least 84\% accuracy, although the optimal prompting strategy varied across models. In general, prompting the model to say ``don't know'' when it is unsure (the uncertainty prompt variant) improves accuracy, particularly for GPT-4o and Claude 3.5 Sonnet.

\begin{sidewaystable}[htbp]
\begin{threeparttable}
\caption{Evaluation results across models and prompting strategies. The gold standard was derived from a tractography atlas (section \ref{sec:eval}). Results in \textbf{bold} show the best performing model for each prompt, and \underline{underlined} indicates the best performing prompt for a given model. Mean $\pm$ std accuracies across four repeats.}
\label{table:prompt_accuracy}
\begin{tabular}{r|cc|cc|cc}
\toprule
\multicolumn{1}{l}{} &
  \multicolumn{2}{c|}{minimal} &
  \multicolumn{2}{c|}{reasoning} &
  \multicolumn{2}{c}{chain-of-thought} \\
model &
  standard &
  UPV &
  standard &
  UPV &
  standard &
  UPV \\ \hline
\multicolumn{1}{c|}{Claude 3.5 Sonnet} &
  0.76 ± 0.06 &
  0.79 ± 0.03 &
  0.80 ± 0.07 &
  \underline{\textbf{0.85 ± 0.03}} &
  0.66 ± 0.02 &
  0.75 ± 0.01 \\
Llama 3 8b &
  0.82 ± 0.01 &
  0.83 ± 0.01 &
  0.57 ± 0.01 &
  \underline{0.84 ± 0.01} &
  0.56 ± 0.01 &
  0.50 ± 0.00 \\
GPT-4o &
  0.78 ± 0.00 &
  \underline{\textbf{0.85 ± 0.02}} &
  0.65 ± 0.02 &
  0.80 ± 0.02 &
  0.68 ± 0.02 &
  0.74 ± 0.04 \\
GPT-4 Turbo &
  \textbf{0.89 ± 0.01} &
  \textbf{0.85 ± 0.04} &
  \textbf{0.82 ± 0.01} &
  0.83 ± 0.02 &
  \textbf{0.89 ± 0.02} &
  \underline{ \textbf{0.91 ± 0.02}} \\
\bottomrule
\end{tabular}
\end{threeparttable}
\end{sidewaystable}

\begin{figure}[htbp]
  \begin{center}
  \includegraphics[width=\textwidth]{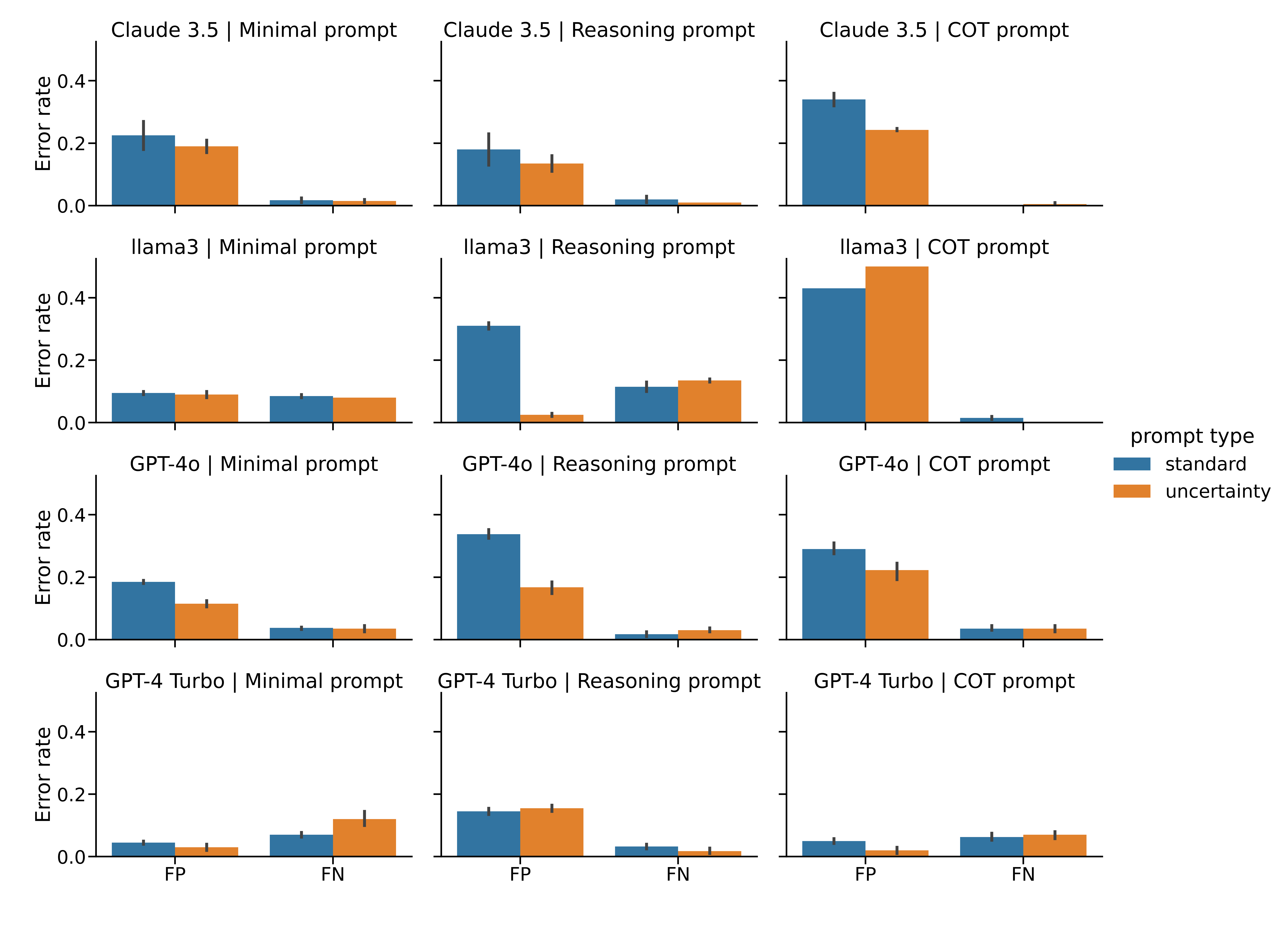}
  \caption{Bar charts showing the false positive and false negative rates across models and prompts. False positives (FP) refer to pairs classified as connected by the LLM that are not connected in the tractography atlas, and false negatives (FN) to those that are connected in the atlas but not classified as such by the LLM. Error bars show the spread across four repeats. The blue and orange bars correspond to the standard and uncertainty variants of the prompts, respectively.}
  \label{fig:error_rates}
  \end{center}
\end{figure}

Figure \ref{fig:error_rates} shows the error rates from the different models. We can see that false positives are the dominant error type, which are reduced by using the uncertainty prompt in most cases. The minimal prompt also performs well across most models. OpenAI's GPT-4 Turbo consistently achieves high accuracy, as it generates fewer false positives than the other models. Llama3 gives the most variable results; it performs fairly well with the minimal prompt, but has very low specificity with the COT prompt. 

We performed an additional stability analysis with ten repeats of the minimal prompt and GPT-4 Turbo. The model produced consistent outputs for 84 out of the 100 region pairs across all repeats, and the overall accuracy was $0.89\pm0.03$ (range: $0.82-0.92$). The order that the regions were presented in the prompt were reversed for half of the repeats. This accounted for much of the variability in the results; the ordering was particularly important for pairs containing subcortical regions and areas of the cingulate cortex. For example, for the region pair (Brainstem, Precentral), which is part of the corticospinal tract, the LLM correctly classified the regions as connected for all five repeats when the Brainstem was first in the prompt, but gave the incorrect response for all repeats when the regions were presented the opposite way round. This could reflect biases in the training data.

\subsubsection{Cost Efficiency}
Average costs for the different prompts and models are shown in Table \ref{tab:cost_efficiency}, calculated based on token costs as of April 24th, 2025. Note that the costs are for querying 100 region pairs (or 50 region pairs, if queried with the regions in both forward and reverse ordering). Llama3 is an open-source model, so it is free to use. The minimal prompt is an order of magnitude cheaper than the reasoning and COT prompts, which is due to the single output token required from the LLM.

\begin{table}[h]
\begin{threeparttable}
\caption{Average cost in USD to retrieve connection estimates for 100 region pairs, using three different prompting approaches.
\label{tab:cost_efficiency}}
\setlength{\tabcolsep}{25pt}
\begin{tabular}{l ccc}
\toprule
        Model & minimal & reasoning & chain-of-thought \\
        \hline
        Llama3 8b & 0.0 $\pm$ 0.0 & 0.0 $\pm$ 0.0 & 0.0 $\pm$ 0.0 \\
        Claude 3.5 Sonnet & 0.01 $\pm$ 0.0 & 0.28 $\pm$ 0.01 & 0.44 $\pm$ 0.02\\
        GPT-4o & 0.01 $\pm$ 0.0 & 0.12 $\pm$ 0.02 & 0.27 $\pm$ 0.02 \\
        GPT-4 Turbo & 0.05 $\pm$ 0.0 & 0.62 $\pm$ 0.04 & 1.06 $\pm$ 0.04 \\
    \bottomrule
    \end{tabular}
\end{threeparttable}
\end{table}

\subsection{LLM uncertainty estimates are a good predictor of error}
\begin{figure}[htbp]
  \begin{center}
  \includegraphics[width=1\textwidth]{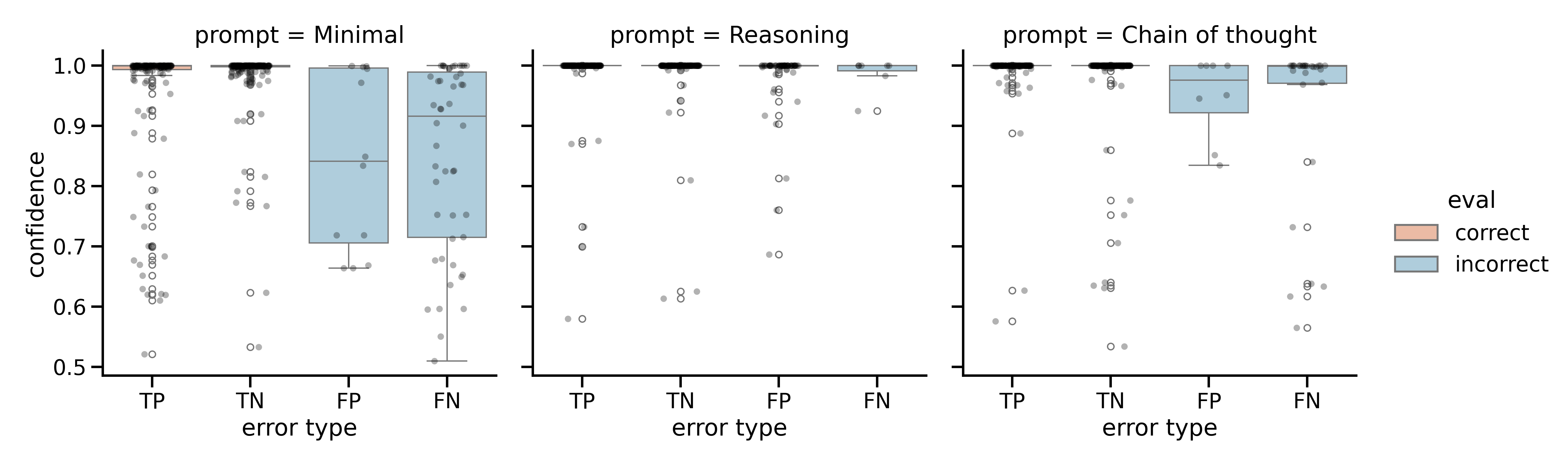}
  \caption{Model confidence for different prompts, derived from the log-probs variable. True positives (TP) and true negatives (TN) are region pairs for which the LLM is in agreement with the tractography atlas. False positives (FP) and false negatives (FN) are region pairs where the LLM diverges from the tractography atlas.}
  \label{fig:model_confidence}
  \end{center}
\end{figure}

Figure \ref{fig:model_confidence} shows the model confidence scores for different prompting strategies. We show results from GPT-4 Turbo with the uncertainty prompt variant, as this was the best performing combination (section \ref{sec:prompt_engineering_results}). Where the model agrees with the tractography atlas, confidence scores tend to be high. The model’s confidence is significantly lower for results that diverge from the tractography atlas across all prompts, tested with a Mann-Whitney U test (minimal prompt: $U = 16935.0, p = 3.4\times10^{-16}$; reasoning prompt: $U = 16300.0, p = 2.2\times10^{-8}$; COT prompt: $U = 10672.5, p = 4.8\times10^{-10}$). Therefore, low confidence from the LLM on the final True/False token can indicate that its responses are more likely to be inaccurate, which provides a useful uncertainty marker for downstream analysis.

\subsection{Discrepancies between the LLM and tractography atlas}\label{sec:disagreement}
\begin{table}[h!]
\begin{threeparttable}
\caption{Region pairs where the LLM consistently disagreed with the tractography atlas. The connected column indicates whether the regions were connected in the atlas. Results from GPT-4 Turbo with the uncertainty variant of the COT prompt.
\label{table:disagreement}}
\setlength{\tabcolsep}{25pt}
\begin{tabular}{cccc}
\toprule
\textbf{Region 1} & \textbf{Region 2} & \textbf{Connected} & \textbf{Confidence} \\ \hline
LateralOrbitofrontal & SuperiorFrontal & True & 0.747 \\ 
Putamen & SuperiorFrontal & True & 0.989 \\ 
SuperiorFrontal & Thalamus & True & 0.998 \\ 
Brainstem & Precentral & True & 0.999 \\ 
Brainstem & SuperiorFrontal & True & 1.000 \\ 
RostralMiddleFrontal & Thalamus & True & 1.000 \\ 
Parahippocampal & PosteriorCingulate & False & 1.000 \\ 
    \bottomrule
    \end{tabular}
\end{threeparttable}
\end{table}

Table \ref{table:disagreement} lists the region pairs for which there was a consistent disagreement between GPT-4 Turbo and the tractography atlas, for the best performing prompt and model. Most include subcortical regions, which suggests that the LLM is lacking knowledge about projection fibres. Figure \ref{fig:case_studies} highlights two representative examples:

\begin{figure}[htbp]
  \begin{center}
  \includegraphics[width=\textwidth]{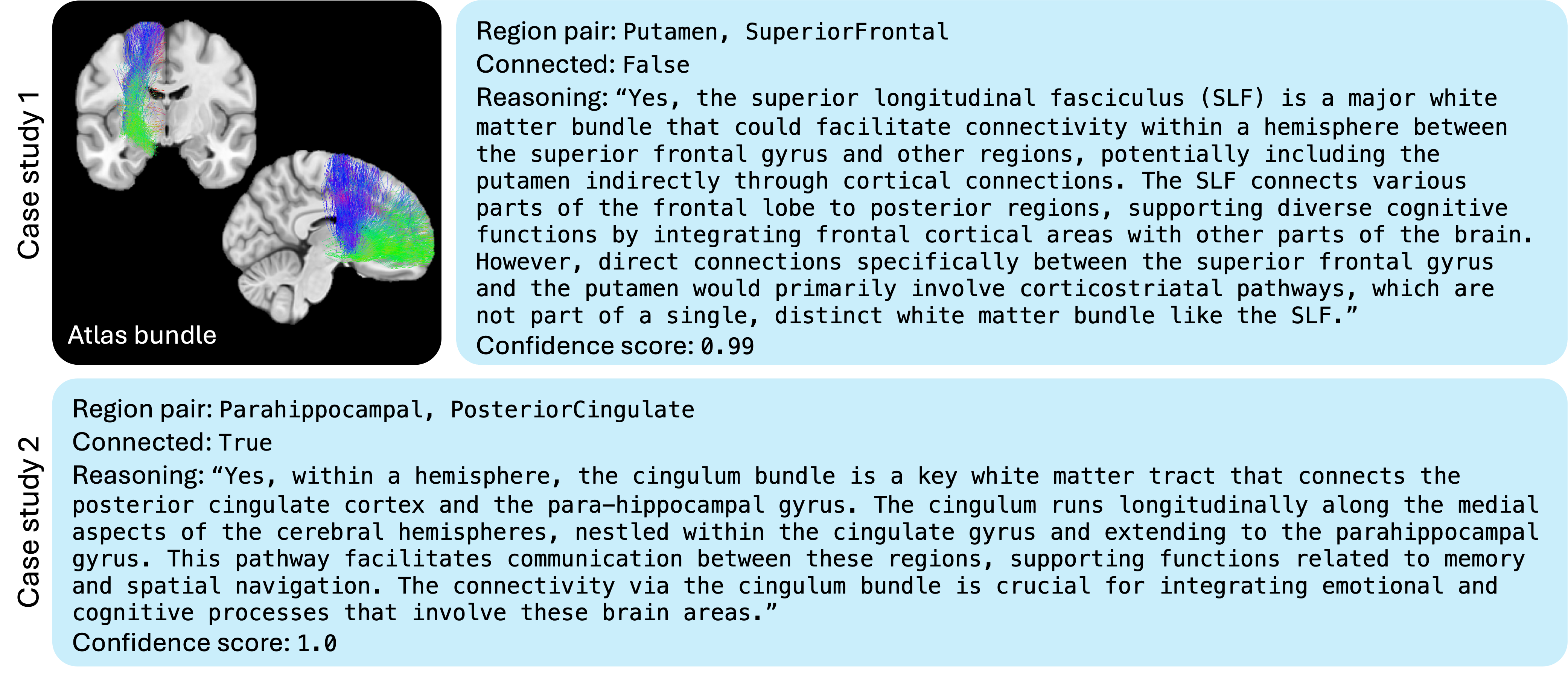}
  \caption{Case study 1, left: Streamlines from the tractography atlas that connect the putamen and the superior frontal region (coronal and sagittal views, left hemisphere). Right: sample output from GPT-4 Turbo. Case study 2: sample output from GPT-4 Turbo for a connection between the parahippocampal gyrus and the posterior cingulate.}
  \label{fig:case_studies}
  \end{center}
\end{figure}

\paragraph{Case study 1: Putamen and Superior Frontal lobe}
The putamen and the superior frontal lobe were strongly connected in the tractography atlas but were not classified as such by the LLM. It is documented that the putamen receives input from Brodmann area 6, which extends to the superior frontal lobe~\citep{nieuwenhuys_human_2008}. The LLM tentatively suggests a connection via the superior longitudinal fasciculus (SLF), but ultimately returns ``False'' for its final output, which is incorrect. Its suggestion of SLF involvement is also incorrect, as the SLF is an association tract that connecting cortical regions~\citep{liang_comparative_2022}. The phrasing of the prompt may have contributed to this error: the corticostriatal pathways would not typically be described as a single white matter ``bundle'', which may be why the LLM disregards them as a possible connection. This highlights the challenges in selecting appropriate terminology to use in the prompt.

\paragraph{Case study 2: Parahippocampal Gyrus and Posterior Cingulate}
These regions were not connected in the tractography atlas, but were consistently and confidently identified as being connected by the LLM. This connection corresponds to the parahippocampal cingulum, which is a well documented fibre bundle~\citep{bubb_cingulum_2018}. Therefore, it is likely a true connection that is missing from the tractography atlas. 

\subsection{RAG can improve accuracy when LLM lacks relevant knowledge}

\begin{table}[h!]
\begin{center}
\begin{threeparttable}
\caption{Accuracy on the HCP-MMP evaluation set, with and without RAG to provide region context. Mean $\pm$ std accuracies across four repeats.
\label{tab:RAG_background}}
\setlength{\tabcolsep}{25pt}
\begin{tabular}{ccc}
\toprule
  & COT prompt & minimal prompt \\
        \hline
        original & $0.72 \pm 0.03$ & $0.73 \pm 0.04$ \\
        RAG & $\mathbf{0.80 \pm 0.04}$ & $0.70 \pm 0.02$ \\
    \bottomrule
    \end{tabular}
\end{threeparttable}
\end{center}
\end{table}

We used RAG to provide contextual information about the HCP-MMP, since the parcellation contains many novel regions. Using RAG increased the accuracy in the evaluation set from 72\% to 80\% with the COT prompt, indicating that the model is able to make more accurate classifications when provided with contextual information. However, when used with the minimal prompt, the RAG pipeline led to an increase in false positives that outweighed the increase in correctly identified connections. This shows that careful prompt engineering is required when incorporating external information, to avoid hallucinations.

\subsection{RAG can provide citations for identified connections}\label{sec:RAG_pubmed_results}
\begin{figure}[htbp]
  \begin{center}
  \includegraphics[width=0.8\textwidth]{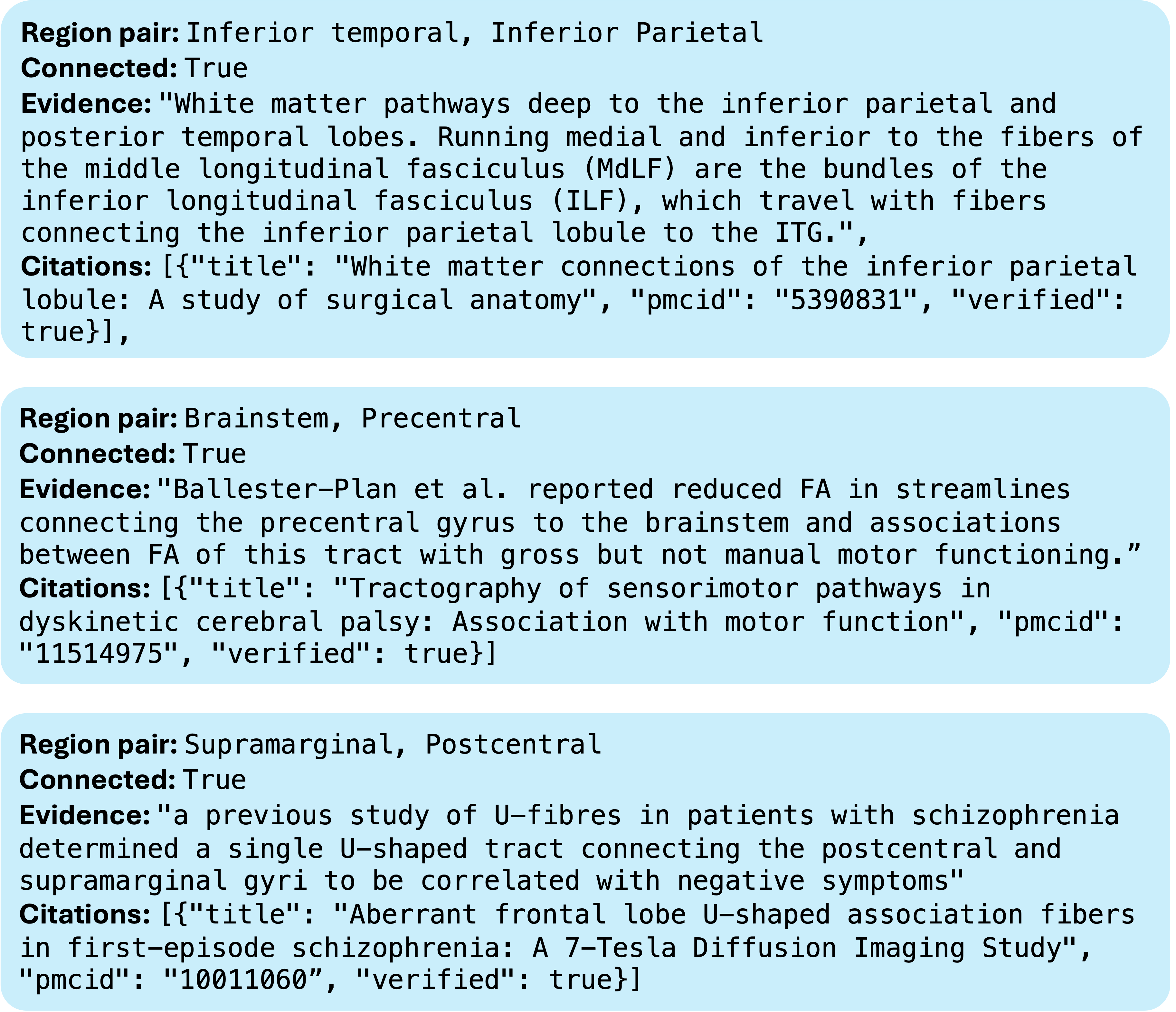}
  \caption{Example outputs from the RAG pipeline. All the region pairs shown were classified as connected in the tractography atlas.}
  \label{fig:RAG_examples}
  \end{center}
\end{figure}

We developed a RAG pipeline that can provide verifiable citations for connections, based on a database of neuroscience papers. We tested the pipeline on the DKT evaluation set using GPT-4 Turbo, and it yielded an accuracy of $80.5 \pm 1.2\%$ across four repeats. This is lower than the best accuracy achieved with GPT-4 Turbo using internal knowledge alone. However, the advantage of using RAG is that verifiable context for the connections are obtained, with citations from the literature. Some examples are shown in Figure \ref{fig:RAG_examples}. All of the region pairs shown are strongly connected in the white matter atlas. The RAG pipeline is able to provide extracts from published papers that support these connections, alongside the titles and PMCIDs of the articles. Across all trials, the PMCIDs and titles returned by the LLM were consistent with the entries in the database, indicating that there were no hallucinations of fabricated publications.

\subsection{LLM-augmented filtering can improve accuracy of connectomes for pathology spreading models}
\begin{figure}[htbp]
  \begin{center}
  \textbf{Model fit for different connectome filtering methods}\par\medskip
    \includegraphics[width=0.45\textwidth]{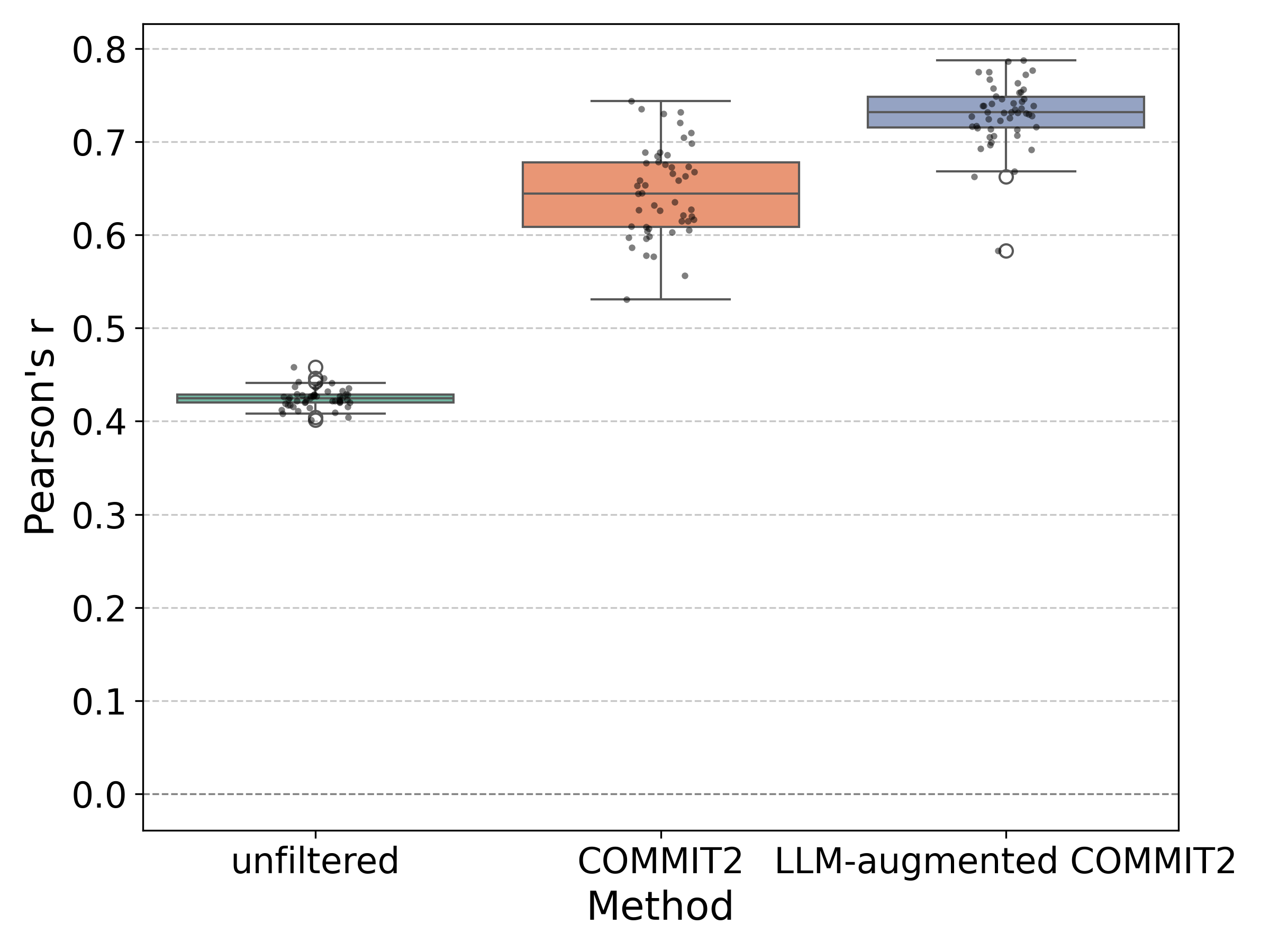}
    \label{fig:subfig1}
    \includegraphics[width=0.45\textwidth]{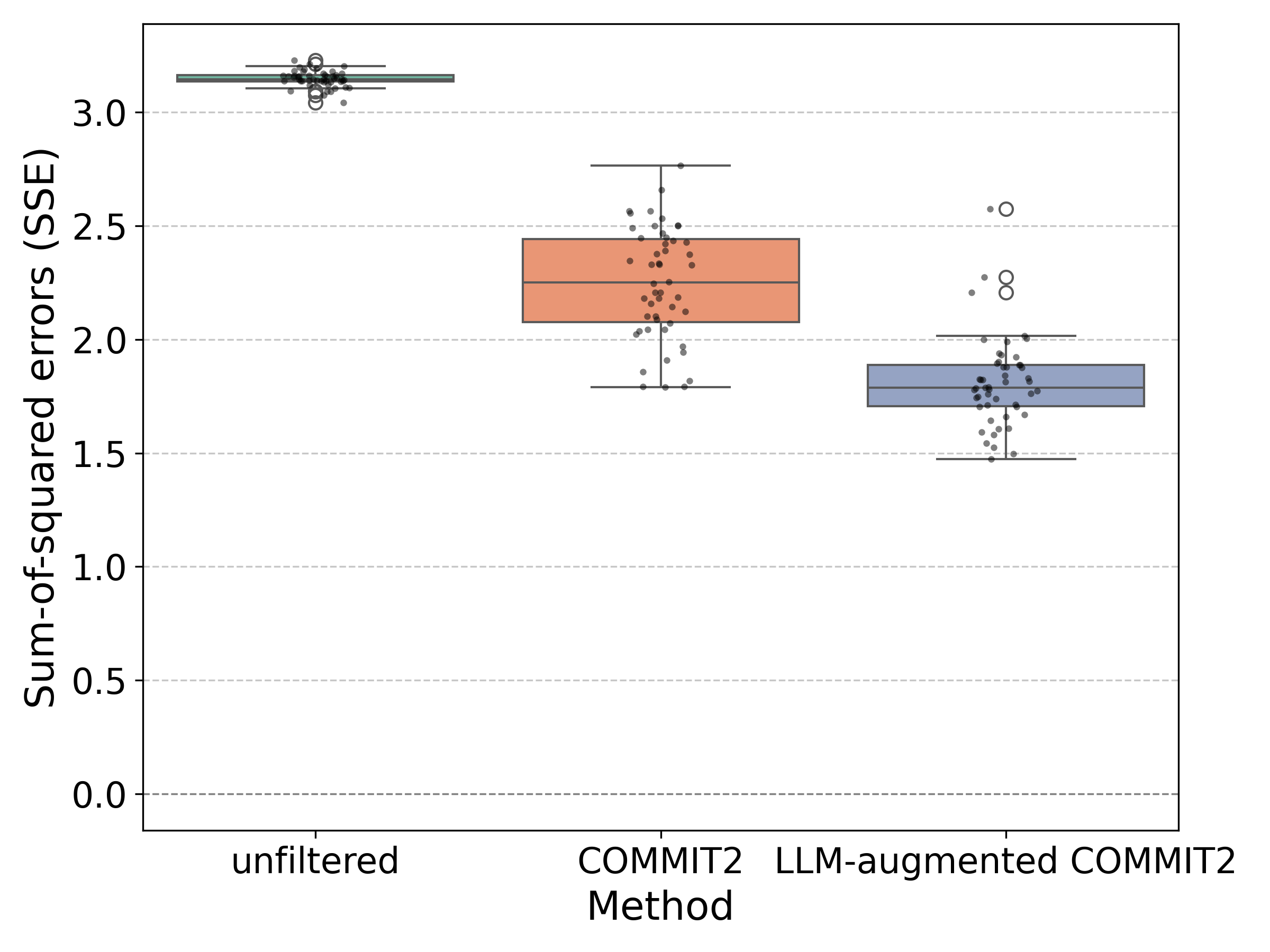}
    \label{fig:subfig2}
  \caption{Comparing the fit of the network diffusion model using connectome substrates with three different levels of filtering: no filtering,  filtering with COMMIT2, LLM-augmented COMMIT2 filtering. Each data point corresponds to the model output for a connectome from a different HCP participant, with the same target data. Left: Pearson's r between model estimates and average tau-PET pattern. Right: SSE between the model estimates and measured data.}
  \label{fig:filtering_model_fit}
  \end{center}
\end{figure}

We used a pathology spreading model to test whether incorporating LLM-derived priors into the connectome filtering process could provide an improved characterisation of tau deposition in Alzheimer's Disease. We propose that more accuracy in the connectome will improve the fit of the network diffusion model to tau-PET data from individuals on the Alzheimer's Disease continuum. Figure \ref{fig:filtering_model_fit} demonstrates a clear improvement in model fit to the data when using LLM-augmented filtering, compared to both COMMIT2 filtered and unfiltered connectomes. Correlation of the model predictions and the measured data is higher with the LLM-augmented filtering, and the SSE is significantly reduced compared to unfiltered connectomes and COMMIT2 filtering alone ($p < 0.001$, repeated measures ANOVA with Bonferroni-corrected paired t-tests). Furthermore, this improvement was significant in comparison to a null distribution of COMMIT2 filtered connectomes with a random selection of connections retained to match the number selected by the LLM, with $p_{permutation} < 0.05$ for SSE and r in all subjects.

\begin{figure}[htbp]
  \begin{center}
  \includegraphics[width=\textwidth]{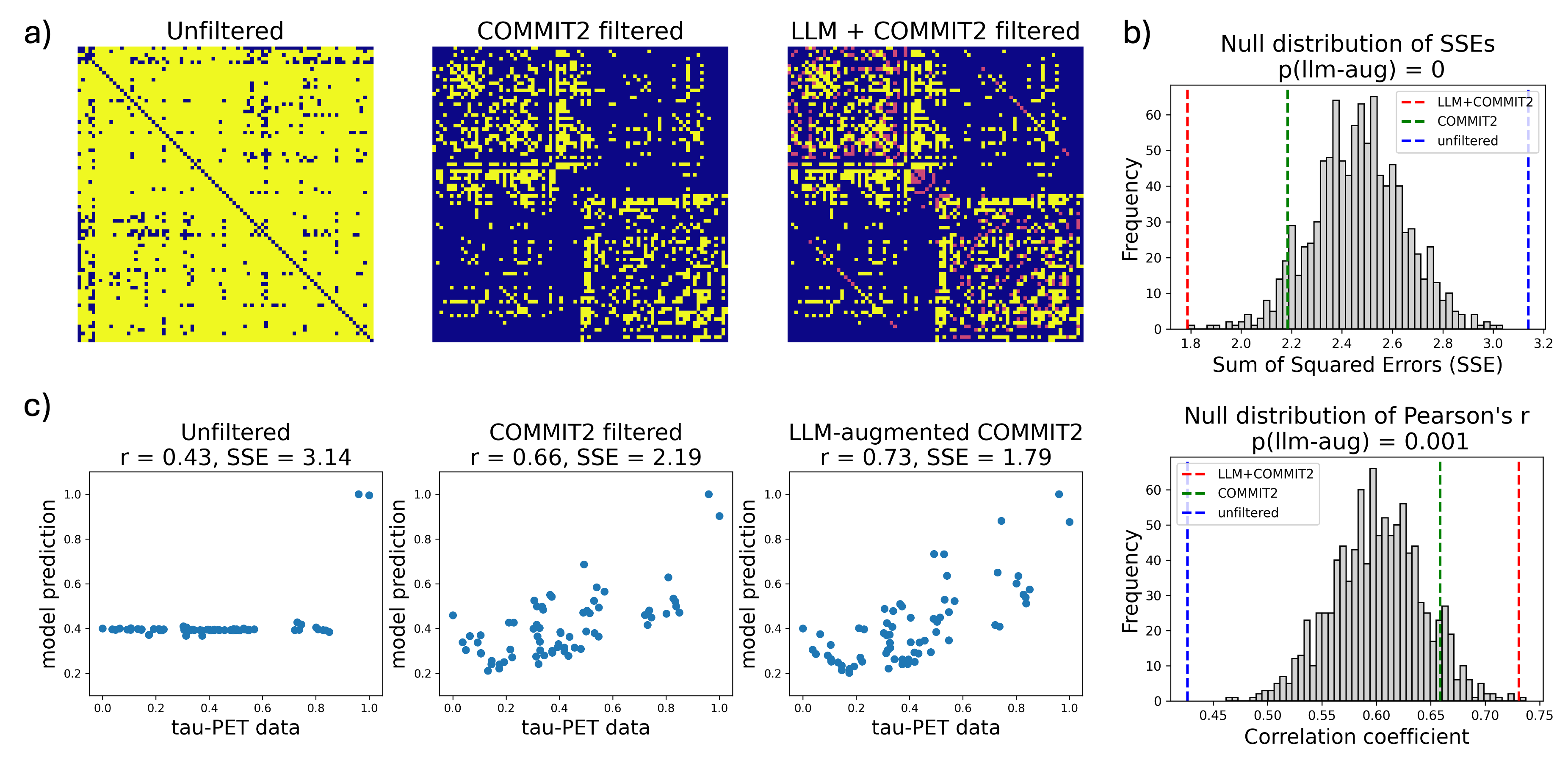}
  \caption{Filtering results from a representative HCP subject. a) Heatmaps of the binarised raw and filtered connectomes, with connections in yellow and non-connected regions in blue. Additional connections that were preserved by LLM-augmented filtering compared to COMMIT2 are shown in pink. b) Comparison of model performance for different connectome substrates, as measured by SSE (top) and Pearson's r (bottom). For each metric, the combined LLM-augmented filtering approach leads to an improvement in model accuracy that exceeds that of a null distribution generated by randomly adding connections to the COMMIT2-filtered connectome. c) Scatter plots of the model prediction against normalised tau-PET SUVRs for each of the connectomes.}
  \label{fig:filtering_example}
  \end{center}
\end{figure}

Figure \ref{fig:filtering_example} shows example results from a representative subject. We can see that the retention of relatively few additional connections compared to COMMIT2 filtering (panel a) leads to a significant improvement in the model's ability to characterise patterns of tau deposition, both in terms of increased correlation between the model prediction and the measured data and a reduction in SSE (panel b and c). Model output saturates for the unfiltered connectome due to the high density of connections. When random connections were added to the COMMIT2 filtered connectome, the model performance tended to decrease, whereas the connections chosen by the LLM lead to improved model accuracy (panel b). This adds evidence that the connections chosen by the LLM are biologically meaningful.

\section{Discussion}
We have examined whether large language models can accurately characterise the existence of white matter pathways, to provide anatomical priors for connectomics. First, we aimed to find which prompting strategy and model gave the most accurate results. We found that the best-performing combinations were able to classify around 90\% of the evaluation set correctly. We also found that the LLM is least confident in its answers where it disagreed with the tractography atlas (Figure \ref{fig:model_confidence}). This indicates that the model's confidence in its answer can reflect the accuracy of its response, providing useful uncertainty quantification for downstream tasks, although it could be a circular effect in that much of the LLM's knowledge will have come from tractography studies. Interestingly, a minimal prompt provided similarly accurate results to a chain-of-thought prompt for some of the models, which runs contrary to previous results showing that additional reasoning steps improve accuracy in LLMs~\citep{wei_chain--thought_2023}. This could be due to the binary classification nature of the problem, where additional reasoning steps can cause an increase in false positives as the models have more opportunities for hallucinations. The minimal prompt also provides a good balance between accuracy and cost, as demonstrated in Table \ref{tab:cost_efficiency}. Another advantage of the minimal prompt is that its uncertainty scores were lower for incorrect answers, compared to the COT and reasoning prompts (Figure \ref{fig:model_confidence}), indicating that additional reasoning steps cause the model to become more confident, even in answers that are incorrect.

We also explore different methods of incorporating RAG into the pipeline, to augment the LLM's knowledge and provide citable references for its outputs. We have shown that RAG can improve the accuracy of the LLM's outputs by providing additional domain knowledge, and that it is possible to use a database of neuroscience papers to provide citations for the LLM's responses. However, we were not able to attain as high accuracy with this approach compared to using the LLM's internal knowledge alone. It could be that performance is limited by the size of the database, as only a relatively small proportion of PubMed papers are open-access. Future work could incorporate a more highly curated selection of texts, and use web-search capabilities and vision models to expand the scope of the database to internet articles and diagrams. Furthermore, a more exhaustive comparison of parameters such as chunk size and retrieval mechanism may yield further improvements. 

Finally, we demonstrated an application of the LLM-derived priors, using them in combination with microstructure-based filtering to retain likely true positive connections. We show that the LLM-augmented filtering approach leads to improved accuracy in a network spreading model of Alzheimer's pathology, compared to regular filtering or unfiltered connectomes. This approach demonstrates the kind of use cases for the LLM that we envision to be most useful, where knowledge from the LLM is used in conjunction with imaging features to improve connectome accuracy. The training data for the LLMs are inherently asymmetric, in that they are skewed towards positive examples, descriptions of a bundle connecting regions A and B, rather than negative examples citing explicit absence of a connection. Therefore, the filtering strategy relies on LLM-derived priors to identify connections that are likely to exist and should therefore be retained. We deliberately avoid using LLM outputs to remove connections, as we consider them to be more reliable when expressing strong confidence in the presence of a connection than in its absence. 

\subsection{Limitations and future work}

One key area of future work is to construct a better evaluation set. The current evaluation set is based on a tractography atlas, and so will suffer from the biases that affect tractography~\citep{maier-hein_challenge_2017}. We attempted to mitigate this by focusing on the most strongly connected regions, assuming that these are more likely valid connections than those containing fewer streamlines. However, in section \ref{sec:disagreement}, we show an example where LLM identified a connection that is missing from the atlas, despite evidence for its existence in the literature. Promisingly, this suggests that our framework might be useful for highlighting connections difficult to track with tractography, but it also highlights inaccuracies in the evaluation set. This might be improved in future by collaborating with neuroanatomists to validate the set of connected and unconnected regions. Such inaccuracies may also contribute to some apparent false positives in the RAG experiment from section \ref{sec:RAG_pubmed_results}, for which the LLM found supporting evidence in the literature. Ultimately, the lack of gold standard connectome-level templates remains a challenge for the field, but one that will hopefully be resolved by future large-scale initiatives~\citep{descoteaux_millennium_2025}.

Another direction for future work could be to build upon the existing RAG pipeline to tune the LLM-derived priors for a specific population, for example reflecting changes in the connectome that would be expected with ageing or between different disease groups. The output from the current framework doesn't reflect inter-individual variability, which limits its applicability to normative connectomes. However, we see the LLM as providing a set of constraints or priors for imaging-based connectomics, rather than acting as a stand-alone tool, as demonstrated in the filtering approach. In this way, the imaging-based features provide the subject-specific information, but the LLM's knowledge is used to guide connectome construction according to known features of anatomy that are shared across the population.

In future, vision models could also be integrated into the RAG pipeline (Figure \ref{fig:RAG_parcellation}), which would enable us to incorporate diagrams and provide the models with visual information relating to the parcellation. The current framework requires a parcellation with anatomically meaningful region names, which are not always available (e.g. the Schaefer atlas~\citep{schaefer_local-global_2018}). Providing visual representations of the regions in the parcellation may remove the need for meaningful region names and provide additional relevant context to the models.

Although we have demonstrated that LLMs can provide accurate information about brain connections, we found cases where the models hallucinate plausible-sounding connections (e.g. case study 1, Figure \ref{fig:case_studies}). In general, the most common errors are false positives rather than false negatives (Figure \ref{fig:error_rates} and Table \ref{table:disagreement}), indicating that the LLMs are inclined to suggest connections existing between brain regions that were not present in our atlas, rather than missing existing connections. The propensity of LLMs to hallucinate is well-documented \citep{huang_survey_2023}. We showed that prompting the models to admit uncertainty can reduce false positives from hallucinations (Figure \ref{fig:error_rates}), although other strategies such as prompt ensembles could improve accuracy further~\citep{pitis_boosted_2023}. Since the models are closed-source we are unable to check the veracity of the training data, so errors in the training literature could be propagated into the model outputs. Fine-tuning an open-source model on a curated literature database of findings from gold-standard methods would be one route to improve traceability of the model's knowledge base, but would require significant computational resources to be competitive with the state-of-the-art proprietary models~\citep{kaplan_scaling_2020}. Hallucinations may also arise as a result of the LLMs' post-training phase, in which they are evaluated on test scores that penalise uncertainty and reward confident answers~\citep{kalai_why_2025}. Post-training procedures that promote honest expressions of model confidence would promote greater accuracy in scientific applications such as ours.

The outputs from the framework can be used to filter implausible streamlines and highlight likely missing connections from tractography, based on existing anatomical knowledge. Integrating data-driven insights with LLM predictions has shown promise in the construction of knowledge graphs~\citep{abdulaal_causal_2023}. We could use a similar approach by iteratively updating a tractogram based on LLM-derived priors, guided by uncertainty quantification in both data sources~\citep{schroder_false-positive_2021,behrens_characterization_2003}. Global tractography and reinforcement learning tractography are other areas where LLM-derived priors could be used~\citep{theberge_track--learn_2021,jbabdi_bayesian_2007}. This would provide complementary information to imaging-derived features, to promote biologically-plausible streamlines and ultimately improve the accuracy and interpretability of tractography.

\section{Conclusion}
We have shown that large language models can provide a novel route for injecting prior neuroanatomical knowledge into connectomics studies, with demonstrated benefits for improving the sensitivity of tractography filtering. We also show how external sources of information, such as scientific papers, can be incorporated into the framework to provide additional verification.

\section*{Data and Code Availability}

Code and RAG database will be made available upon peer review.

The ORG Fiber Clustering White Matter Atlas is available at \href{https://github.com/SlicerDMRI/ORG-Atlases}{https://github.com/SlicerDMRI/ORG-Atlases}.

Data from the ADNI are publicly and freely available from the Laboratory of Neuro Imaging (LONI) Image and Data Archive upon sending a request that includes the proposed analysis and the named lead investigator, at \href{https://adni.loni.usc.edu/data-samples/adni-data}{https://adni.loni.usc.edu/data-samples/adni-data}.

\section*{Author Contributions}

Elinor Thompson: Conceptualisation, Methodology, Software, Validation, Formal Analysis, Investigation, Writing - Original Draft, Visualisation. Tiantian He: Conceptualisation, Writing - Review \& Editing. Anna Schroder: Writing - Review \& Editing. Ahmed Abdulaal: Conceptualisation. Alec Sargood: Writing - Review \& Editing. Sonja Soskic: Writing - Review \& Editing. Henry F.J. Tregidgo: Writing - Review \& Editing. Daniel C. Alexander: Conceptualisation, Resources, Writing - Review \& Editing, Supervision, Project administration, Funding acquisition.

%\section*{Funding}

%Funding text (optional).

\section*{Declaration of Competing Interests}

The authors have no competing interests to declare that are relevant to the content of this article.

\section*{Acknowledgements}

E.T, T.H, A.S, A.A, A.S, S.S are supported by the Wellcome Trust (221915). A.S, T.H and A.A are also supported by the EPSRC-funded UCL Centre for Doctoral Training in Intelligent, Integrated Imaging in Healthcare (i4health) (EP/S021930/1). D.C.A and H.F.J.T are supported by the Wellcome Trust (221915), MRC grant MR/W031566/1 and the NIHR UCLH Biomedical Research Centre.

This work was supported by compute credits from a Cohere For AI Research Grant and the OpenAI Researcher Access Program. These grants are designed to support academic partners conducting research with the goal of releasing scientific artifacts and data for good projects.

Data collection and sharing for the Alzheimer's Disease Neuroimaging Initiative (ADNI) is funded by the National Institute on Aging (National Institutes of Health Grant U19AG024904). The grantee organization is the Northern California Institute for Research and Education. In the past, ADNI has also received funding from the National Institute of Biomedical Imaging and Bioengineering, the Canadian Institutes of Health Research, and private sector contributions through the Foundation for the National Institutes of Health (FNIH) including generous contributions from the following: AbbVie, Alzheimer’s Association; Alzheimer’s Drug Discovery Foundation; Araclon Biotech; BioClinica, Inc.; Biogen; Bristol- Myers Squibb Company; CereSpir, Inc.; Cogstate; Eisai Inc.; Elan Pharmaceuticals, Inc.; Eli Lilly and Company; EuroImmun; F. Hoffmann-La Roche Ltd and its affiliated company Genentech, Inc.; Fujirebio; GE Healthcare; IXICO Ltd.; Janssen Alzheimer Immunotherapy Research \& Development, LLC.; Johnson \& Johnson Pharmaceutical Research \& Development LLC.; Lumosity; Lundbeck; Merck \& Co., Inc.; Meso Scale Diagnostics, LLC.; NeuroRx Research; Neurotrack Technologies; Novartis Pharmaceuticals Corporation; Pfizer Inc.; Piramal Imaging; Servier; Takeda Pharmaceutical Company; and Transition Therapeutics.

%\section*{Supplementary Material}

%Supplementary Material (created during production as a web link to online material).

\printbibliography

\appendix

\section{Appendix}\label{sec:Appendix}

\subsection{Prompt for RAG}
\paragraph{System prompt:} You will be presented with the names of a pair of brain regions and some snippets from the scientific literature. 

\noindent Your task is to determine if the brain regions are connected by white matter fibers in humans, based on the text you are shown. 

\noindent Approach the task step-by-step: Identify which snippets come from HUMAN studies. Discard any snippets that mention animals. Are the relevant brain regions described in the text? Does the text provide evidence for a white matter connection between the two regions? 

\noindent If there is relevant information in the text, extract a json with keys as follows: 
    \begin{itemize}
        \item connection: allowable values are `True', `False' or `don't know'
        \item evidence: contains any quotes from the text that support your assertion.
        \item citations: a list of dictionaries, each with `title': title of the referenced paper, and `pmcid': the pubmed central ID of the relevant snippet
    \end{itemize}

\noindent Return your answer as a json object with the keys `connection', `evidence', and `citations'

\paragraph{User Prompt:}
Are these brain regions connected by white matter fibres within a hemisphere: \{region1\}, \{region2\}? The connection can be in either direction. 

\noindent Text snippets: \verb|<context> \{context text\} </context>|.

\noindent Remember you are looking for a white matter connection between \{region1\} and \{region2\}.
\noindent Return your answer as a JSON object with the keys `connection', `evidence' and `citations'. Return a JSON only and no other text.

\end{document}